\documentclass[superscriptaddress,prl,twocolumn,amsmath,twocolumn,amssymb,nofootinbib]{revtex4}
\usepackage{graphicx}
\usepackage{color}
\usepackage{eurosym}
\usepackage{dcolumn}
\usepackage{bm}
\usepackage[final]{pdfpages}
\usepackage{setspace}
\usepackage{pdfpages}
\graphicspath{{./images/}}
\usepackage{subfig}
\usepackage{amsmath}

\newcommand{\superscript}[1]{\ensuremath{^{\textrm{#1}}}}
\newcommand{\subscript}[1]{\ensuremath{_{\textrm{#1}}}}

\begin{document}
\title{On Chip Manipulation of Single Photons from a Diamond Defect}

\date{\today}

\author{J. E. Kennard}
\affiliation{Centre for Quantum Photonics, H. H. Wills Physics Laboratory \& Department of Electrical and Electronic Engineering, University of Bristol, Merchant Venturers Building, Woodland Road, Bristol, BS8 1UB, UK}
\affiliation{National Physical Laboratory, Hampton Road, Teddington, Middlesex, TW11 0LW, UK}
\author{J. P. Hadden}
\author{L. Marseglia}
\affiliation{Centre for Quantum Photonics, H. H. Wills Physics Laboratory \& Department of Electrical and Electronic Engineering, University of Bristol, Merchant Venturers Building, Woodland Road, Bristol, BS8 1UB, UK}
\author{I. Aharonovich}
\affiliation{School of Physics and Advanced Materials, University of Technology, Sydney, P.O. Box 123, Broadway, New South Wales 2007, Australia}
\author{S. Castelletto}
\affiliation{School of Physics, University of Melbourne, Parkville, Victoria 3010, Australia}
\affiliation{Presently at ARC Centre of Excellence for Engineered Quantum Systems, Physics and Astronomy Department, Macquarie University, North Ryde, Sydney, Australia}
\author{B. R. Patton}
\affiliation{Centre for Quantum Photonics, H. H. Wills Physics Laboratory \& Department of Electrical and Electronic Engineering, University of Bristol, Merchant Venturers Building, Woodland Road, Bristol, BS8 1UB, UK}
\author{A. Politi}
\affiliation{Centre for Quantum Photonics, H. H. Wills Physics Laboratory \& Department of Electrical and Electronic Engineering, University of Bristol, Merchant Venturers Building, Woodland Road, Bristol, BS8 1UB, UK}
\affiliation{Presently at Center for Spintronics and Quantum Computation, University of California Santa Barbara, USA}
\author{J. C. F. Matthews}
\affiliation{Centre for Quantum Photonics, H. H. Wills Physics Laboratory \& Department of Electrical and Electronic Engineering, University of Bristol, Merchant Venturers Building, Woodland Road, Bristol, BS8 1UB, UK}
\author{A. G. Sinclair}
\affiliation{National Physical Laboratory, Hampton Road, Teddington, Middlesex, TW11 0LW, UK}
\author{B. C. Gibson}
\author{S. Prawer}
\affiliation{School of Physics, University of Melbourne, Parkville, Victoria 3010, Australia}
\author{J. G. Rarity}
\author{J. L. O'Brien}
\affiliation{Centre for Quantum Photonics, H. H. Wills Physics Laboratory \& Department of Electrical and Electronic Engineering, University of Bristol, Merchant Venturers Building, Woodland Road, Bristol, BS8 1UB, UK}

\begin{abstract}
Operating reconfigurable quantum circuits with single photon sources is a key goal of photonic quantum information science and technology. 
We use an integrated waveguide device comprising of directional couplers and a reconfigurable thermal phase controller to manipulate single photons emitted from a chromium related colour centre in diamond. Observation of both a wave-like interference pattern and particle-like sub-Poissionian autocorrelation functions demonstrates coherent manipulation of single photons emitted from the chromium related centre and verifies wave particle duality. 
\end{abstract}

\maketitle

Linear optical circuits (LOCs) are a powerful platform for performing  quantum optical experiments \cite{jwp-interfrev-2012-revmodphys}. With the addition of high-fidelity photon sources, feed-forward and high-fidelity photon detection, LOCs form a core component of proposed quantum enhanced technologies \cite{kn-nat-409-46,ob-natphot-3-687} and are a leading platform for performing fundamental quantum experiments \cite{jwp-interfrev-2012-revmodphys}. Recently, LOCs have begun to be integrated into monolithic waveguide circuit architectures \cite{po-sci-320-646,Peruzzo2012, Shadbolt2011}. With inherent interferometric stability and increasing component miniaturisation \cite{Bonneau2012}, integrated LOCs offer the prospect of performing increasingly complex and flexible quantum optical tasks \cite{pe-sci-329-1500,Shadbolt2011}, particularly in conjunction with high efficiency on-chip photodetection \cite{Sprengers, Pernice2012}. In previous experiments the photons used in LOCs have been generated by spontaneous parametric sources of photon pairs and subsequent heralding of single photons contaminated by higher photon number terms. Conversely single defect centres in diamond emit via single electron transitions and therefore produce true single photons, even at saturation, making them an attractive scalable photon source for LOCs.


Here we report the coupling of a room temperature colour centre single photon source and an integrated LOC. The single photons are emitted from a chromium related diamond colour centre which operates at room temperature \cite{Aharonovich2010}. The LOC is a two path interferometer comprised of two directional couplers and an electrically tuned phase shifter on one arm. The LOC enables on-chip manipulation of quantum information encoded onto the path of a verifiable single photon, and contains all of the components required to build arbitrary unitary manipulations. We verify wave-particle duality by utilising wave-like self interference and particle-like detection statistics observed simultaneously with this single device. Here we include both directional couplers and active phase shifters\footnote{We note that single photons from nitrogen vacancy centres have been coupled to a straight GaP waveguide\cite{Fu2008}}, which opens up the possibility of using true single photons in complex LOCs, for example studying single photon modal entanglement\cite{Papp2009} on a scale of complexity only practically achievable with integrated optics. 

SPDC has been a particularly useful source of photons for numerous LOC demonstrations. However, the spontaneous emission statistics of
parametric down conversion means these sources must be operated at low occupation probability to suppress multi-photon terms. Experiments are performed
non-deterministically relying on post-selection of photons in coincidence. Without the multiplexing of multiple down conversion processes---with heralding detectors, fast switches and optical delay \cite{mi-pra-66-053805,ma-arxiv-1007.4798}---SPDC cannot scale efficiently for many quantum photonics applications. Alternative photon sources, such as quantum dots \cite{Kako2006}, single atoms
\cite{Kuhn2002}, ions \cite{Keller2004} and dye molecules \cite{Hwang2011}, based on quantised energy transitions have been employed as single photon sources (SPSs), but require either cryogenic temperatures (dots) or a vacuum environment(atoms, ions) to ensure the requisite identical photons.

\begin{figure*}[t]

\subfloat[]{\label{fig:fig1a}\includegraphics[width=0.5\textwidth]{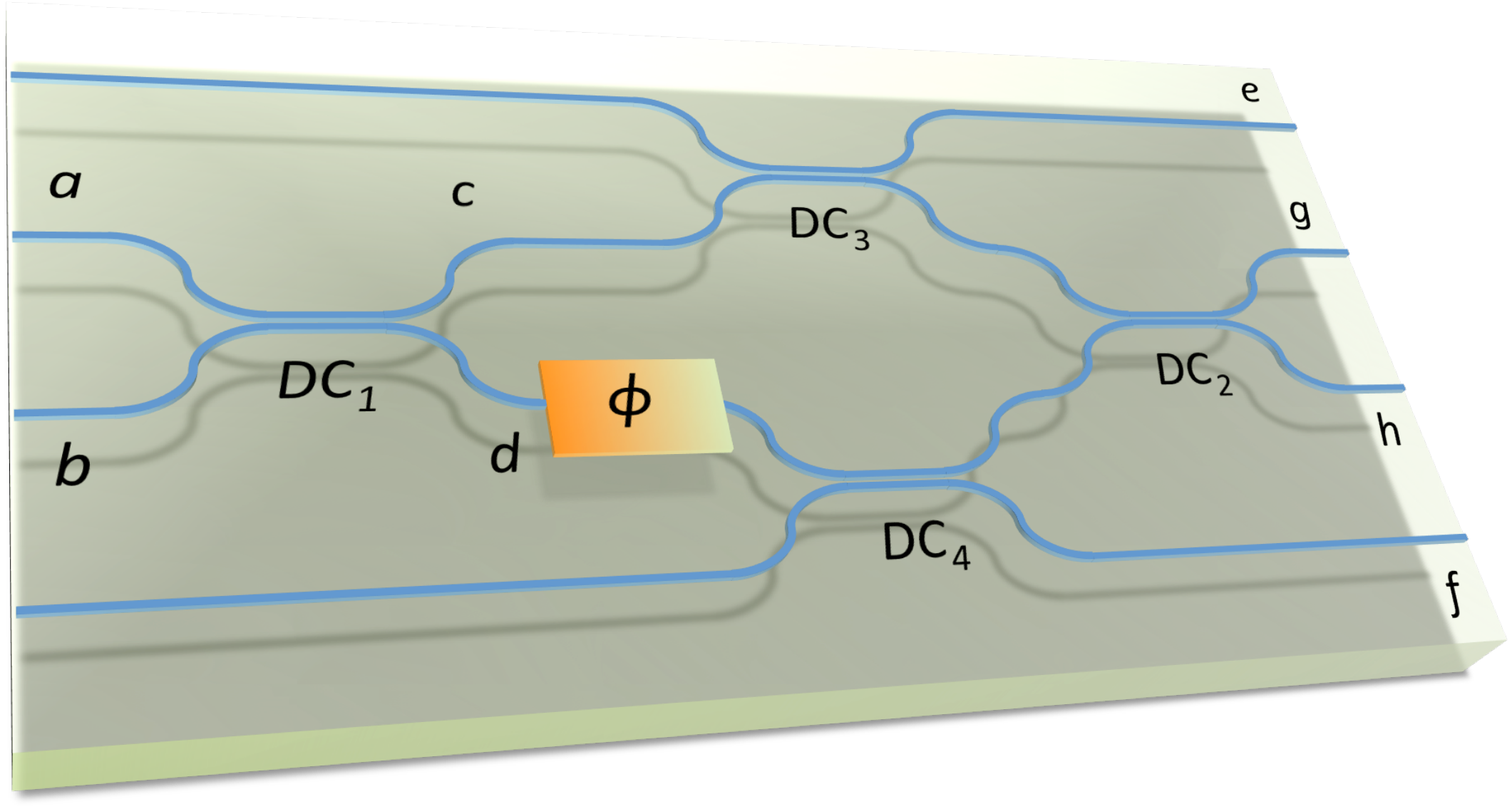}}
\subfloat[]{\label{fig:fig1b}\includegraphics[width=0.35\textwidth]{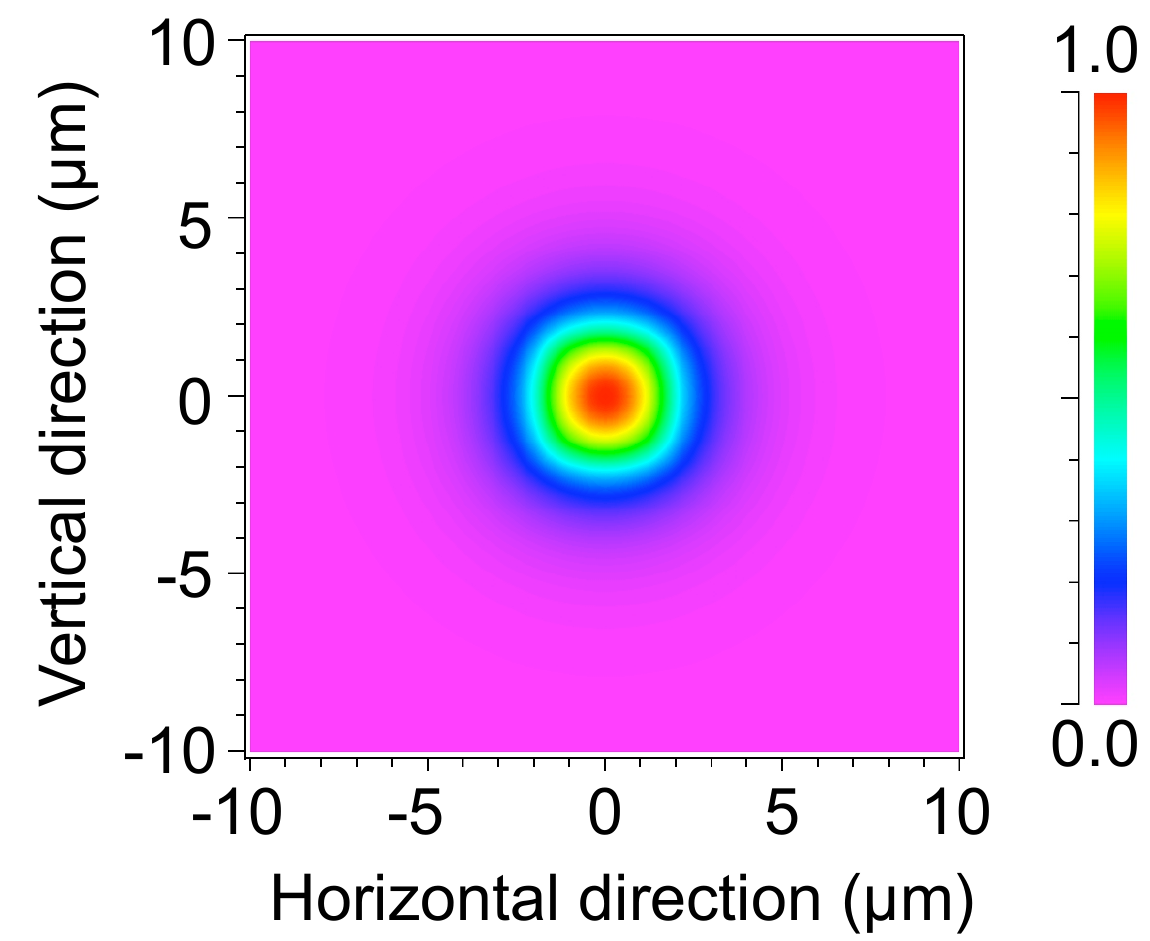}}

\caption{The silica on silicon waveguide circuit.
(a) Silica on silicon waveguide circuit. Directional couplers DC\subscript{1} and DC\subscript{2} have reflectivity = 1/2 and DC\subscript{3} and DC\subscript{4} have reflectivity = 1/3. A resistive heater at waveguide {\it d} allows a relative phase {$\phi$} to be imparted. Photons from an individual chromium related centre are coupled into waveguide {\it a} from polarisation maintaining fibre (PMF) and are sent to APDs via PMF from waveguides {\it e - f}. (b) Simulated mode of 780 nm light in chip.}
\end{figure*}

Individual diamond defect centres display atom-like optical spectra, with sub-10 nm ZPL linewidths \cite{Aharonovich2011a} even at room temperature. Thus they are convenient sources of single photons requiring nether trapping nor in principle cryogenic operation, ensuring they are excellent candidates for room temperature sources. Emission linewidths are still orders of magnitude away from being lifetime limited which limits room temperature operation of centres to single photon applications such as quantum key distribution (QKD) \cite{Beveratos2002}. Traditionally, this distinguishability may be overcome by reducing phonon interactions by placing the colour centre in a cryogenic environment \cite{Bernien2012,Sipahigil2011} however by reducing the available emission density of states through coupling the emitter to a high Q cavity lifetime limited linewidths could be achieved \cite{Kaupp2013} at or near room temperature.

\begin{figure*}[t]
\centering

\subfloat[]{\label{fig:fig2a}\includegraphics[width=0.5\textwidth]{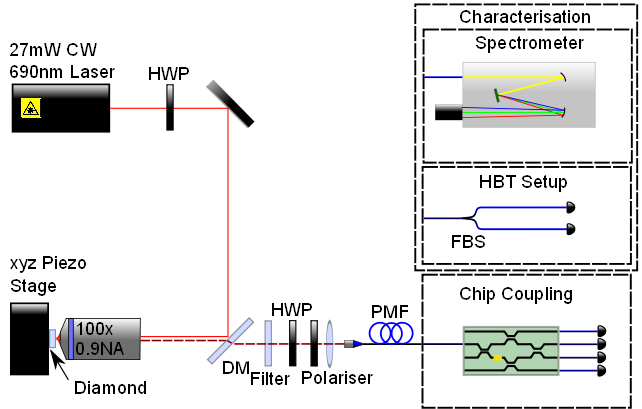}}
\subfloat[]{\label{fig:fig2b}\includegraphics[width=0.5\textwidth]{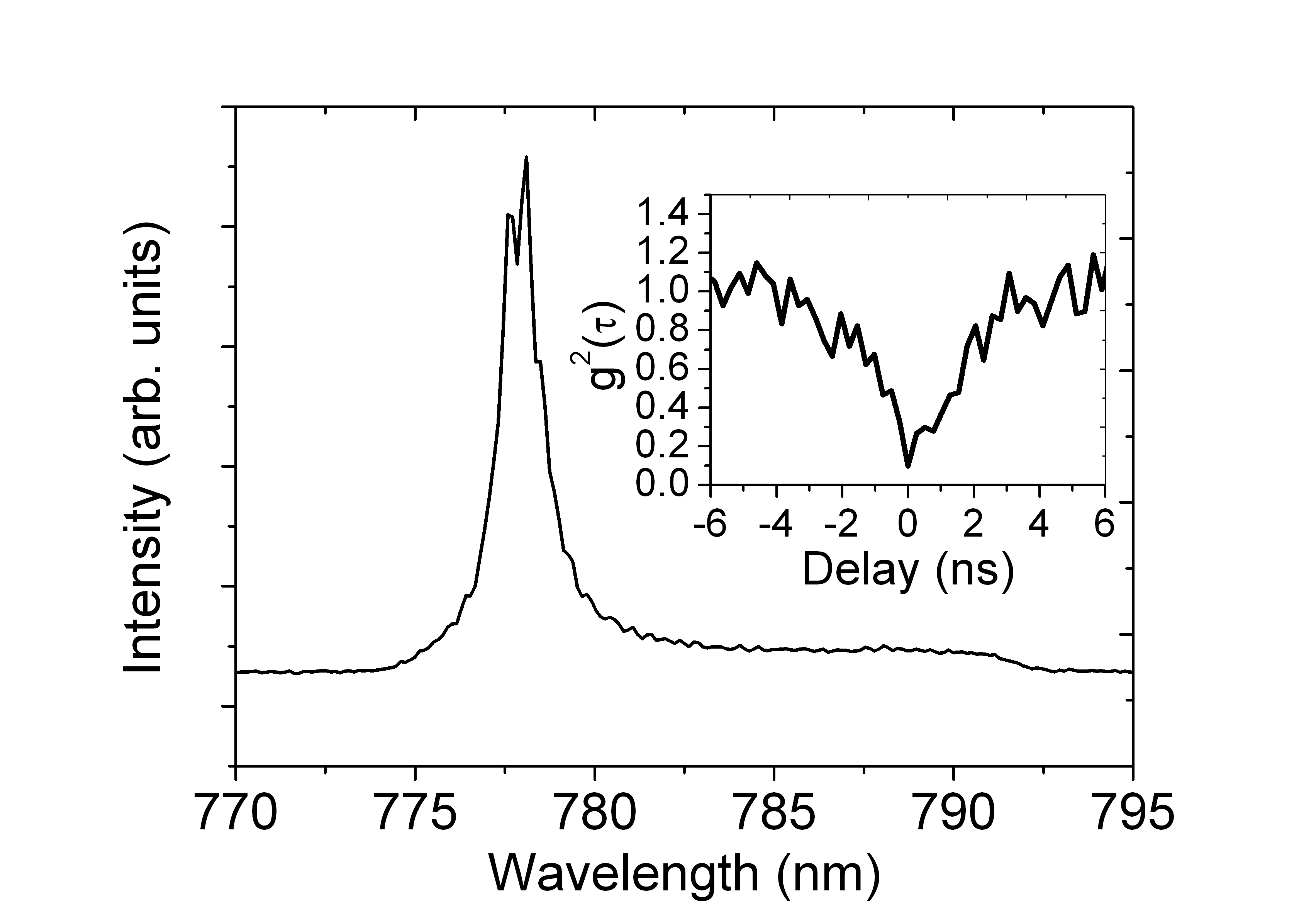}}

\caption{Experimental setup and characterisation of chromium related centres. (a) Schematic diagram of the optical setup used in this paper. Photons collected from chromium related centres being excited with a 27 mW 690 nm continuous wave laser are fibre coupled and subsequently sent to one of three measurements - a spectrometer, a Hanbury Brown-Twiss correlation setup or the chip described in the main text. (b) The emission spectrum of a chromium related centre used in the waveguide experiments. Inset: Second order auto-correlation function of the chromium related centre.}
\end{figure*}


Here we employ chromium related defect centres, which are particularly bright \cite{Aharonovich2010b}, emitting the majority of photons into the ZPL \cite{Aharonovich2010}, which facilitates the creation of a high efficiency SPS. The centre is created by co-implantation of chromium and oxygen ions using 50 keV and 19 keV energies, respectively into Type IIa diamond ($<$1 ppm nitrogen and $<$1 ppb boron) \cite{Aharonovich}. The sample is annealed at 900 C for an hour. Some individual centres do exhibit blinking or photo bleaching behaviour, although this seems to be a symptom of the implantation process, given that chromium related centres in nanodiamond do not \cite{Aharonovich2009}. Emission line widths of 4 nm, polarised emission and excited state lifetimes of the order of a nanosecond mean the centres are attractive SPSs for quantum photonic technologies. Due to inhomogeneous broadening ZPLs are found within the range 730 nm - 790 nm ensuring easy integration with existing silica on silicon waveguides, allowing centres to be chosen that emit a wavelength that is inherently low loss. Unlike the negative nitrogen vacancy centre(NV), which has two orthogonal dipoles, the chromium related defect's ZPL consists of only a single dipole transition, ensuring all emitted photons have the same linear polarisation at room temperature, allowing coherent manipulation within LOCs as waveguide modes are typically polarisation dependant. These factors ensure photons emitted from chromium related defects are suitable for observation of single particle quantum behaviour in a LOCs, with the prospect of multi-photon LOC applications with the use of cavity QED.

We report the operation of a diamond colour centre single photon source in conjunction with a monolithic LOC waveguide chip. Single photons emitted from a chromium-related colour centre are guided through the silica-on-silicon circuit \cite{po-sci-320-646,ma-prl-107-163602} shown in figure \ref{fig:fig1a}, enabling manipulation of a path encoded qubit and observation of wave and particle effects simultaneously within the single device.  The waveguides are lithographically fabricated from silica doped with boron and germanium oxides to control refractive index, on a silicon substrate. The dopant levels in the core and cladding are controlled to define a refractive index contrast of $\Delta = 0.5\%$; together with a 3.5 $\mu$m $\times$ 3.5 $\mu$m dimension core, the waveguides support only the fundamental mode of near infra-red light (in the region of 780nm light for the purposes of our demonstration, figure.~\ref{fig:fig1b} shows the simulated mode for 780 nm light). The waveguide circuit comprises four inputs and four outputs, and four directional couplers, {\it DC\subscript{1-4}}, each with reflectivity $\eta_i$ and modelled with $2\times 2$ transition matrices

\begin{align} \label{eq:DC}
DC_i = \left( \begin{array}{cc}
\sqrt{\eta_i} & i \sqrt{1-\eta_i} \\
i \sqrt{1-\eta_i} & \sqrt{\eta_i} \\
\end{array} \right)     
\end{align}
The circuit is designed to have reflectivities $\eta_{1}=\eta_{2} = 1/2$ and $\eta_{3}=\eta_{4} = 1/3$. The 
basis of the circuit is an interferometer formed from $DC_1$ and $DC_2$, in which the internal phase is controlled via voltage applied across a thermal phase shifter $\phi$, fabricated by lithographically patterning a metal layer in the form of a resistive heating element directly above the waveguide and connecting contact pads used to apply voltage (not shown).  It is straightforward to compute a $4\times 4$ transition matrix 
that models the circuit ($U_\textrm{chip}$), which shows
the single photon entering optical mode $a$ of the circuit ideally evolves according to 

\begin{figure*}[t]
\centering

\subfloat[]{\label{fig:fig3a}\includegraphics[width=0.45\textwidth]{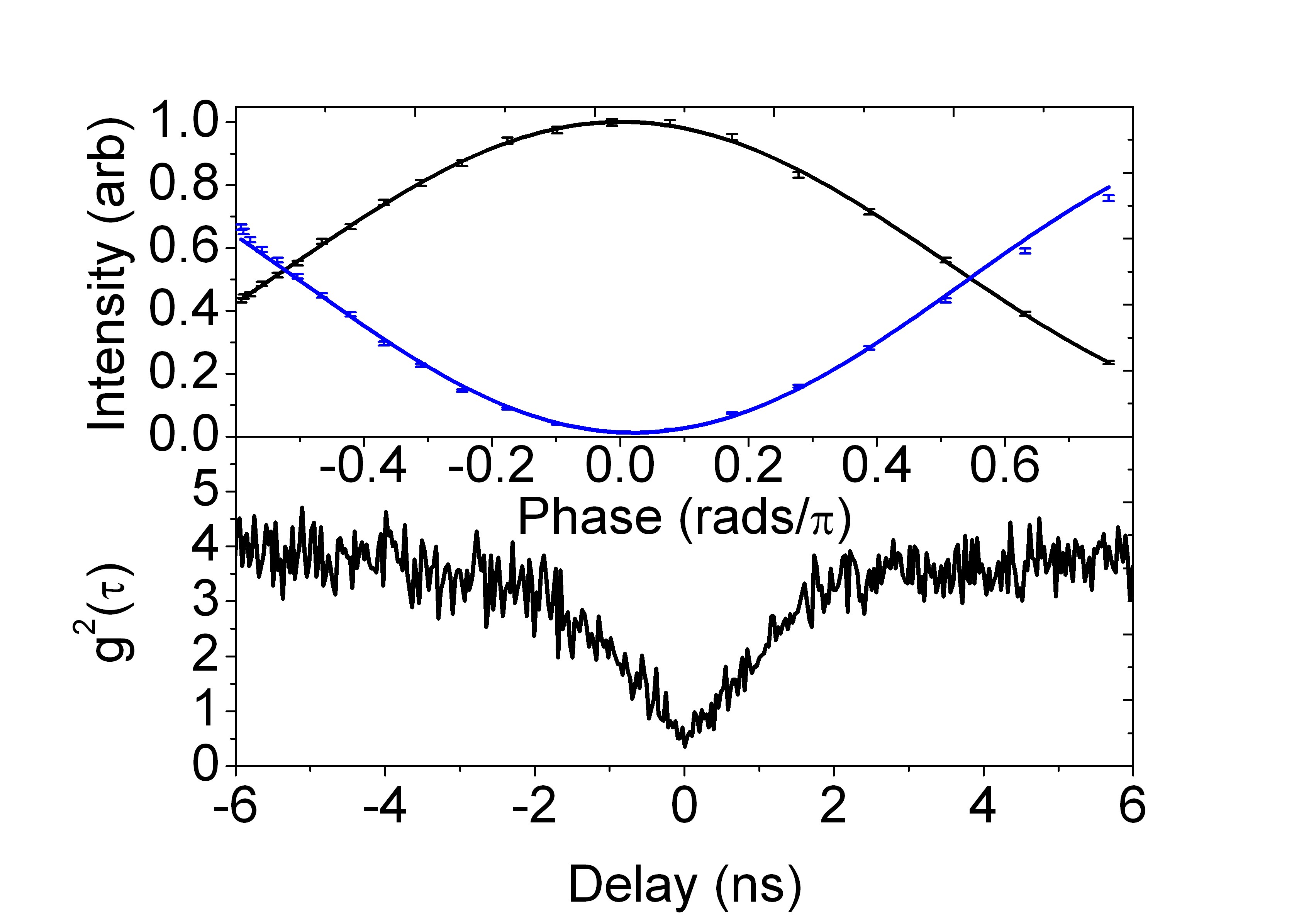}}
\subfloat[]{\label{fig:fig3b}\includegraphics[width=0.45\textwidth]{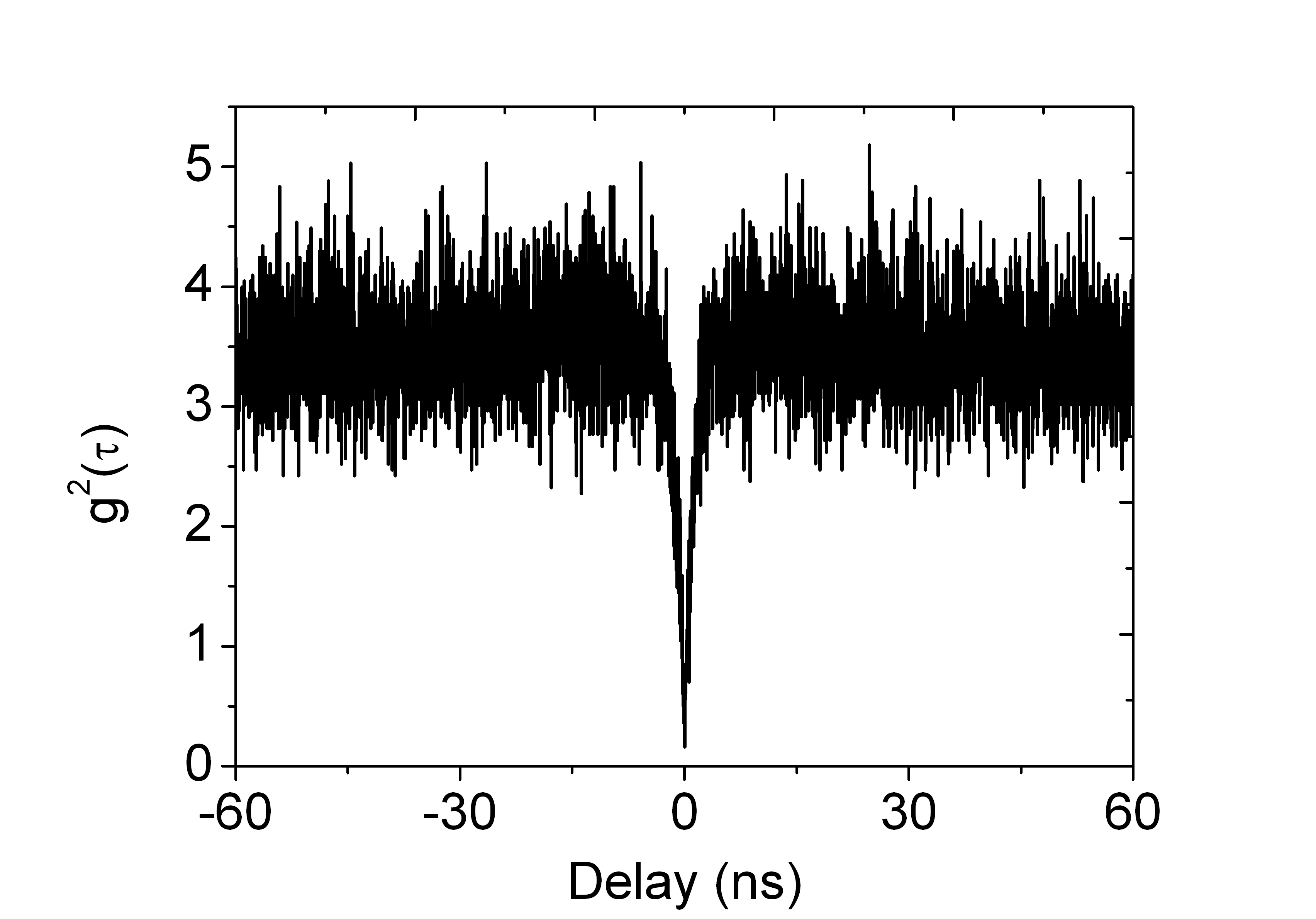}}

\caption{Experimental data acquired from photons coupled to the integrated waveguide circuit. (a) Above: Intensity measured from waveguides {\it g} and {\it h} as a function of $\phi$. The upper (black) line is from waveguide {\it h} and the lower (blue) line is from waveguide {\it g}. Below: Second order auto-correlation function of the chromium related centre measured from outputs {\it e} and {\it f}. (b) Second order auto-correlation function of the chromium related centre measured from outputs {\it h} and {\it f} when $\phi = 0$.}
\end{figure*}

\begin{eqnarray}\label{eq:chip}
a_a^\dagger \! \stackrel{U_{\textrm{chip}}}{\longrightarrow} \!\frac{i}{\sqrt{3}} \!
\left(\!
a_e^\dagger 
\!-\! e^{i \frac{\phi}{2}} \sin{\frac{\phi}{2}} a_g^\dagger
\!+\!e^{i \frac{\phi}{2}} \cos{\frac{\phi}{2}} a_h^\dagger 
\!+\!i e^{i \phi } a_f^\dagger
\!\right)
\label{eq:chip}
\end{eqnarray}
Thus the probability to detect photons at output $g$ and $h$ has a sinusoidal phase dependence, enabling
wave-like interference fringes to be observed by measuring the rate of photons detected in each arm as a function of phase.


The verification of particle-like behaviour is achieved by measuring second order correlation statistics between the output modes. The second order correlation function between two modes is given by $g^{(2)}(\tau) = \left\langle I_1(t)I_2(t+\tau)\right\rangle/\left\langle I_1(t)I_2(t)\right\rangle$.  This is obtained experimentally by measuring time intervals between photons detected in each of the two arms after a directional coupler. A single photon entering optical mode $a$ is detected at each of the output modes with probability given by the modulus square of the amplitudes given in Eq. \ref{eq:chip}. It takes finite time for a single photon emitter to emit each photon, therefore the measurement of correlations between the two arms are strongly suppressed for zero time differences.  This effect is known as anti-bunching.  Since both modes are coupled to the same single photon source, through a directional coupler, this is equivalent to a Hanbury Brown-Twiss style measurement of the second order auto correlation function. Measurement of $g^{(2)}(0) = 0$ shows that the photons measured are truly emitted in single events.


Centres were addressed optically with a laser scanning fluorescence confocal microscope shown in figure \ref{fig:fig2a}. Off resonant excitation was achieved via a linearly polarised 27 mW 690 nm continuous wave diode laser and focused onto the sample with a 0.9 NA microscope objective. The sample was mounted onto an 3-axis piezoelectric stage, allowing both precise location of and  stable collection from chromium related centres. Both fluorescence and reflected excitation was collected with the same objective with the fluorescence subsequently transmitted through a 700 nm dichroic mirror. A bandpass filter in the range 770 nm - 790 nm enabled rejection of the first and second order Raman scattering with a half wave plate and linear polariser aligned vertically for output polarisation control. The fluorescence was collected into a single mode 5.0 $\mu$m core polarisation maintaining fibre, acting as the confocal aperture. Typical count rates when measuring fluorescence from a single chromium related centre with silicon APDs were 0.1x10\superscript{6} counts per second. Additionally, this light could then be passed to various experiments for analysis. The photon statistics of centres was determined by measuring the second order auto correlation function between two chosen detectors. Figure \ref{fig:fig2b} shows the $g^{(2)}(\tau)$ function for a 778 nm centre with its corresponding emission spectrum. The dip at zero delay is characteristic of single photon emission, with $g^{(2)}(0)=0.1$. The deviation from zero is attributed to the convolution of signal with the 500 ps jitter on the detectors, which is comparable to the full width half maximum of the $g^{(2)}(\tau)$, $\sim$ 4 ns.

Single photons from the chromium related centres were fibre butt-coupled to the reconfigurable waveguide circuit, detailed above.  The outputs of waveguides $e-h$ were fibre butt-coupled and connected to four silicon APDs for measurement.  Transmission through the waveguide was typically observed to be $\sim$60\%.  To verify the operation of the waveguide circuit, photons from an individual 778 nm centre were coupled into waveguide $a$. By varying the phase $\phi$ and measuring the intensity of outputs $g$ and $h$, an interference fringe, as described in equation \ref{eq:chip}, was observed with visibility $V = (I_{max}-I_{min})/(I_{max}+I_{min}) = 0.971 \pm 0.001$. The full fringe is reproduced in the upper part of figure \ref{fig:fig3a}.  Single photon operation was verified by performing a $g^{(2)}(\tau)$ measurement on the detected signal from modes $e$ and $f$. These equal intensity outputs allowed access to the phase independent part of the modal superposition caused by the 1/2 reflectivity directional coupler DC\subscript{1}.  The lower part of figure \ref{fig:fig3a} shows a $g^{(2)}(0)=0.1$, demonstrating that photons from the chromium related centre maintain their particle statistics after transmission through the LOC.  The non-zero $g^{(2)}(0)$ is again explained by the convolution of the signal and detector jitter.

This verifies the expected behaviour that single photons within our waveguide circuit show wave-like interference behaviour in ports $g$ and $h$ and particle like anti-bunching behaviour in ports $e$ and $f$.  However, to demonstrate duality, one would like to demonstrate simultaneous wave and particle effects behaviour in the same photons.  Whilst performing a second order correlation of the outputs $g$ and $h$ and varying $\phi$ may seem a particularly intuitive method, a $g^{(2)}(\tau)$ may never be obtained when the system is displaying the strongest wave-like behaviour - full constructive/destructive interference, when $\phi = n\pi$.  To investigate duality at this point, a $g^{(2)}(\tau)$ was measured from modes $h$ and $f$ when $\phi = 0$.  At this point, since destructive interference suppresses photon detection in mode $g$, any photons detected at $h$ must have previously displayed wave-like behaviour. Figure \ref{fig:fig3b} shows that the operation displays clear antibunching, with $g^{(2)}(0) = 0.1$, indicating that the photons detected in $h$ must have demonstrated wave-like properties in the interferometer, whilst subsequently being demonstrated to be individual photons by the Hanbury Brown-Twiss experiment. Consequently, the photons must have displayed a wave-like phenomenon within the interferometer and particle-like after DC\subscript{2}, thereby verifying the principles of both complementarity and duality.  Previous experiments have demonstrated wave-particle duality by switching apparatus between an interferometer and a Hanbury Brown-Twiss setup, or placing them in series and observing an interference fringe and an averaged $g^{(2)}(\tau)$ simultaneously\cite{Grangier1986,Braig2003,Aichele2005}. Additional verification has been displayed by the implementation of a delayed choice experiment\cite{Jacques2008}. Our experiment extends this by ensuring that wave particle duality is verified with maximum wave interference visible. 

We have operated a simple LOC circuit with single photons from a chromium related diamond defect centre. Despite some individual centres lacking photostability, the high emission rates, linearly polarised emission and low degree of phonon coupling of the chromium related defect make it an excellent candidate SPS for LOCs. ZPL emission in the near infra red that matches the high transmission and detection efficiencies of current waveguides and detectors technologies is also a key advantage. The high fidelity control of single photons reported here demonstrates the feasibility of manipulating single photon states emitted from other diamond colour centres, and the possibility to harness more complex LOCs. As with other diamond defect centres, at room temperature the emission linewidth of the chromium related centre is not lifetime limited, causing the indistinguishability of sequentially emitted photons to be exceedingly low. However, the prospect of cavity enhanced emission could enable the chromium related centre to produce highly indistinguishable photons\cite{Kaupp2013}, making it a strong candidate SPS for LOCs for fundamental physics and quantum technologies.  

The authors are grateful for financial support from DIAMANT, EPSRC and ERC. This work was carried out with the support of the Bristol Centre for Nanoscience and Quantum Information. J.E.K. is supported by a EPSRC Industrial CASE studentship from NPL. J.L.OB. acknowledges a Royal Society Wolfson Merit Award and a Royal Academy of Engineering Chair in Emerging Technologies. J.C.F.M. is supported by a Leverhulme Trust Early Career Fellowship. The authors would like to acknowledge Brett Johnson and Jeff McCallum for useful advice and discussion.

\bibliography{Waveguides_Chromium}

\end{document}